# A Network Inference Method for Large-Scale Unsupervised Identification of Novel Drug-Drug Interactions

**Roger Guimerà[1,2]***, **Marta Sales-Pardo[2]**

1 Institució Catalana de Recerca i Estudis Avançats (ICREA), Barcelona, Catalonia, Spain, 2 Departament d'Enginyeria Química, Universitat Rovira i Virgili, Tarragona, Catalonia, Spain

## Abstract

Characterizing interactions between drugs is important to avoid potentially harmful combinations, to reduce off-target effects of treatments and to fight antibiotic resistant pathogens, among others. Here we present a network inference algorithm to predict uncharacterized drug-drug interactions. Our algorithm takes, as its only input, sets of previously reported interactions, and does not require any pharmacological or biochemical information about the drugs, their targets or their mechanisms of action. Because the models we use are abstract, our approach can deal with adverse interactions, synergistic/antagonistic/suppressing interactions, or any other type of drug interaction. We show that our method is able to accurately predict interactions, both in exhaustive pairwise interaction data between small sets of drugs, and in large-scale databases. We also demonstrate that our algorithm can be used efficiently to discover interactions of new drugs as part of the drug discovery process.

**Funding:** This work was supported by a James S. McDonnell Foundation Research Award, Spanish Ministerio de Economía y Comptetitividad (MINECO) Grant FIS2010-18639, European Union Grant PIRG-GA-2010-277166 (to RG), European Union Grant PIRG-GA-2010-268342 (to MSP), and European Union FET Grant 317532 (MULTIPLEX). The funders had no role in study design, data collection and analysis, decision to publish, or preparation of the manuscript.

**Competing Interests:** The authors have declared that no competing interests exist.

* E-mail: roger.guimera@urv.cat

## Introduction

Understanding interactions between drugs is becoming increasingly important. A recent large-scale study of older adults (ages 57–85) in the U.S. found that 29% of them use five or more prescription medications concurrently, and that as many as 4% may be at risk of having a major adverse drug-drug interaction [1]. For this reason, the evaluation of drug interactions is "an integral part of drug development and regulatory review prior to its market approval" [2], and institutions like the FDA put much effort in developing guidelines for *in vitro* and *in vivo* studies, as well as for developing *in silico* models and methods.

Potentially beneficial effects of drug interactions, on the other hand, are equally important. Indeed, some drugs show synergistic effects against their targets, which not only increases the efficacy of treatments but may also improve the selectivity and reduce off-target effects [3]. Antagonistic interactions can be used to study the mechanisms of action of drugs [4], and even suppressing interactions between drugs, in which one drug inhibits the action of the other, have been found to be potentially very relevant in the fight against antibiotic-resistant pathogens [5].

More broadly, it is becoming increasingly clear that drug interactions leading to network effects at a systems level are the norm in pharmacology, rather than the exception [6–11]. According to some, these network effects may even be at the root of the dismal results of attempts to develop single-target drugs, and of the simultaneous decline of drug development productivity [7]. Therefore, network pharmacology is emerging as a new paradigm in drug discovery.

However, despite the conceptual appeal of abstract network approaches to drug development, one may argue that the contributions of network analysis have so far been relatively modest. Indeed, most of these contributions have been related to pointing out network properties that make certain proteins *more likely* to be good targets [8], for example connector versus non-connector enzymes [12,13], or central versus peripheral proteins [11]. These contributions notwithstanding, there is little in the form of actual, concrete, examples where network analysis has resulted in a clear application to the discovery of new drugs or to the study of the effects of existing drugs.

Here we present one such application. In particular, we use the information that is encoded in networks of reported drug interactions to predict uncharacterized interactions. Because the models we use are abstract, our approach can deal with adverse interactions as well as synergistic/antagonistic/suppressing interactions or any other type of drug interaction. We show that our method is able to accurately predict drug interactions, and that it can be used efficiently to discover interactions of new drugs as part of the drug discovery process.

## Results

### A network approach for the inference of unknown drug interactions

For specific drug pairs, interactions can be predicted *in silico* from mechanistic or flux balance models of the pathways and

## Author Summary

Over one in four adults older than 57 in the US take five or more prescriptions at the same time; as many as 4% are at risk of a major adverse drug-drug interaction. Potentially beneficial effects of drug combinations, on the other hand, are also important. For example, combinations of drugs with synergistic effects increase the efficacy of treatments and reduce side effects; and suppressing interactions between drugs, in which one drug inhibits the action of the other, have been found to be effective in the fight against antibiotic-resistant pathogens. With thousands of drugs in the market, and hundreds or thousands being tested and developed, it is clear that we cannot rely only on experimental assays, or even mechanistic pharmacological models, to uncover new interactions. Here we present an algorithm that is able to predict such interactions. Our algorithm is parameter-free, unsupervised, and takes, as its only input, sets of previously reported interactions. We show that our method is able to accurately predict interactions, even in large-scale databases containing thousands of drugs, and that it can be used efficiently to discover interactions of new drugs as part of the drug discovery process.

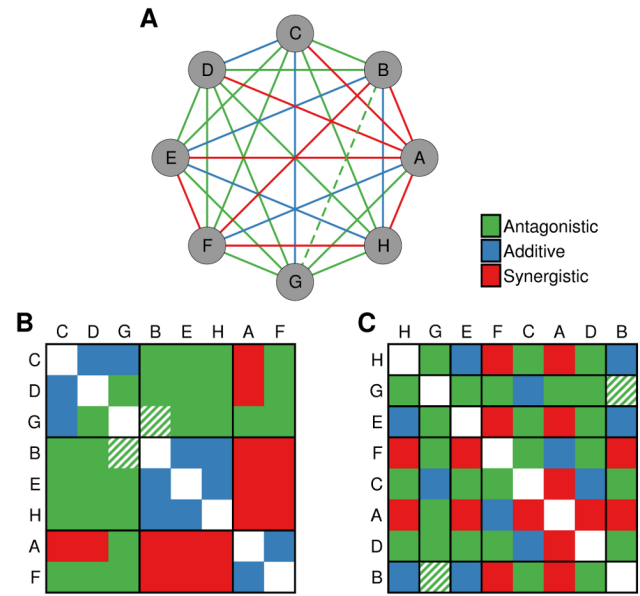

**Figure 1. Stochastic block models for the prediction of unknown drug interactions.** (A) Consider a hypothetical situation in which all of the interactions between drugs $A-H$ are known with the exception of the interaction between $B$ and $G$, which is, in reality, antagonistic. There are many partitions of the drugs into groups. The partition in (B) has high explanatory power (low value of $H(P)$ in Eqs. (5) and (6)), since most drug interactions between a pair of groups are of the same type. Therefore, the predictions of this partition have a large contribution to the estimation of the probability of the unknown interaction. Conversely, the partition depicted in (C) has little explanatory power (high value of $H(P)$) and has a small contribution to the estimation of the probability of the unknown interaction.

doi:10.1371/journal.pcbi.1003374.g001

processes in which their targets are involved [6,14]. However, this approach is difficult to generalize and is, therefore, inappropriate for large-scale identification of interactions and for the identification of interactions between drugs whose mechanisms are not fully understood. Another approach is to use statistical models based on molecular and pharmacological data [15] but, again, such data is not always available. Finally, there are mechanism-independent methods to predict multidrug interactions based on maximum entropy approaches, but these require knowledge of pair interactions [16], which is what we aim to uncover here.

As in other biological problems, network theory [11,17,18] provides a useful, although abstract, alternative to mechanistic and molecular modeling. In a network representation of drug interactions, each node represents a drug and each link represents an interaction between the corresponding pair of drugs. Interactions of different types (for example, synergistic versus antagonistic) are represented by links of different types (Fig. 1A).

Drug interaction networks contain explicit information about the interactions that are known, but also about implicit information about interactions that have never been tested; the question we are concerned with is how to extract this information from the network. Here, we present a network-based approach to predict an interaction $r_{ij}$ between drugs $i$ and $j$ from a network $N^O$ of known drug interactions (which includes $i$ and $j$ but no explicit information about their interaction $r_{ij}$). Our approach deals rigorously with the information contained in the network by means of Bayesian model averaging [19] (Methods). The approach is completely unsupervised and parameter-free.

Within our Bayesian model averaging approach, the only relevant modeling question is what family of models can accurately describe the network of drug interactions. In this regard, it is well established that pairwise drug interactions are largely determined by the cellular functions targeted by the drugs [20–22]. In network terms, this means that the interaction $r_{ij}$ is determined by the cellular functions $\sigma_i$ and $\sigma_j$ of $i$ and $j$, respectively; in other words, nodes can be partitioned into groups (by cellular function) such that the interaction between any pair of nodes depends only on the groups to which they belong (Fig. 1B–C). Stochastic block models are a family of network models that mathematically formalize the idea of group-dependent interactions [23–25]. Although originally proposed in the context of social interactions, stochastic block models are increasingly used to describe the structure of complex networks in general [19,26] and for network inference [19] (Methods). Again, after this choice of plausible models the resulting algorithm is completely unsupervised and parameter-free (Methods).

To benchmark the performance of our algorithm, we consider two alternative heuristic approaches. The first benchmark is based on the idea that similar drugs have similar interactions. In this spirit, we set $r_{ij} = r_{i'j'}$, where $i'$ (respectively, $j'$) is a drug whose known interactions are as similar as possible to those of $i$ ($j$), and $r_{i'j'}$ is a known interaction (Methods). Second, we consider an approach based on the Prism algorithm, which was developed to identify groups of drugs (or genes) with similar interactions to other drugs [20,27]. Instead of averaging over all possible partitions of drugs into groups as done in our Bayesian model averaging approach, we take the partition proposed by Prism and use that partition to make the prediction (Methods).

Additionally, we consider as a baseline the simplest possible algorithm for predicting $r_{ij}$, which is to use the overall rate of each interaction type in the network. For example, if 60% of known interactions in a network are synergistic ($S$) and 40% are antagonistic ($A$), then we set $r_{ij} = S$ with 60% probability and $r_{ij} = A$ with 40% probability. This baseline captures the fact that it is harder to make a prediction when the ratio of $S/A$ interactions is 60/40 than when the ratio is, for example, 95/5.

## Validation on exhaustive pairwise interaction data

We start by testing the algorithms described above against two experiments in which all pairwise interactions between a small set of drugs were exhaustively tested [20,28]. In the first experiment, Yeh and coworkers tested the effect of all pairwise combinations of 21 antibiotics on *E. coli*'s growth [20]. They classified each interaction as synergistic, additive, antagonistic or suppressing. In the second experiment, Cokol and coworkers studied the effect of all pairwise combinations of 13 anti-fungal drugs on the growth of *S. cerevisiae* [28]. They classified interactions as synergistic, additive or antagonistic (except for some interactions that were unresolved).

To study the performance of the algorithms, we simulate situations in which not all pairwise interactions are known. In particular, we simulate a situation in which only a fraction of all interactions are observed, and then try to predict the unobserved interactions (repeated random sub-sampling validation). In each case, we measure the fraction of predictions that are exactly correct (exact classification), as well as the fraction of predictions that deviate from the experimental observation by at most *one level* ($\pm 1$ classification). For example, miss-predicting a synergistic interaction as additive is considered correct by the $\pm 1$ classification metric, but miss-predicting a synergistic as antagonistic or suppressing (or vice versa), or an additive as suppressing (or vice versa) is considered incorrect.

In Fig. 2 we show the results of the validation. As expected, the stochastic block model, the neighbor-based and the Prism-based predictions have accuracies well above the baseline, even when as many as 80% of the interactions are unobserved. In the majority of cases, the stochastic block model is significantly and consistently more accurate than the neighbor-based and the Prism-based predictions with one exception: when the fraction of observed interactions is high ($\geq 70\%$) in the Cokol dataset, in which the neighbor-based prediction is best. Note that as soon as the number of interaction types grows (from 3 in Cokol to 4 in Yeh) or the fraction of observed interactions decreases, the stochastic block model becomes more accurate. Moreover, even when the neighbor-based exact predictions are more accurate, $\pm 1$ predictions are always more accurate for the stochastic block model.

Although the absolute differences of prediction accuracy between the stochastic block model and the neighbor-based approach may seem modest (typically, between 5 and 10 percent points), it is important to note that relative to the baseline the improvements are quite major (Fig. 2E–F). Indeed, when the fraction of observed interactions is 50%, the stochastic block model represents a 29% and a 63% improvement (for the Cokol and Yeh datasets, respectively) in exact classifications over the neighbor-based approach, and a 55% and 66% over the Prism-based approach (always, with respect to the baseline). When we only observe 20% of the interactions, the relative improvements are 126% and 133% over neighbor-based predictions, and 61% and 154% over Prism-based predictions.

## Validation on evolving databases of drug interactions

Next, we test our algorithm against the existence of adverse drug interactions in two drug interaction databases: the database available through the web site Drugs.com and the DrugBank database [29,30]. For the Drugs.com database, we restrict our analysis to major adverse interactions between generic drugs; for the DrugBank, we consider all interactions.

We consider two snapshots of each of the databases. For the Drugs.com database, we collected the first snapshot in May 10, 2010, and the second one in February 22, 2012. A total of 1,518 drugs are listed in both snapshots. There are 32,074 drug interactions present in both instances of the network; $n_N = 1,349$

novel interactions present in the 2012 dataset but not in the 2010 dataset, and $n_S = 165$ spurious interactions present in the 2010 dataset but not present in the 2012 dataset. For the DrugBank dataset, the first snapshot corresponds to January 2009, and the second to April 2012. A total of 1,012 drugs are listed in both snapshots; there are 9,113 drug interactions present in both instances of the network, with $n_N = 1,190$ and $n_S = 233$.

We evaluate to what extent could our network algorithms have predicted which interactions needed to be added to each of the first snapshots (that is, to what extent can the algorithms uncover novel interactions), and which ones needed to be removed (that is, to what extent can they detect spurious interactions). As we show in Fig. 3, the algorithm based on stochastic block models is able to accurately uncover spurious and, especially, novel interactions. In contrast, neighbor-based and Prism-based predictions perform only marginally better than the baseline.

First, we measure the area under the receiver operating characteristic (AUROC) curve (Fig. 3A–B) [31]. In the case of uncovering novel interactions, the AUROC gives the probability that an interaction randomly chosen from those that were added to the first snapshot has a higher score than one randomly chosen from the set of interactions that were never added to the network. For the Drugs.com database, we find this probability to be 0.87 for the stochastic block model, 0.53 for neighbor-based predictions, and 0.52 for Prism-based predictions. For the DrugBank dataset, these probabilities are 0.71, 0.52 and 0.53, respectively.

Similarly, when dealing with spurious interactions, the AUROC gives the probability that an interaction randomly chosen from those that were removed from the 2010 snapshot has a lower score than one randomly chosen from the set of interactions that were not removed from the network. For the Drugs.com database, we find this probability to be 0.73 for the stochastic block model, 0.51 for neighbor-based predictions, and 0.45 for Prism-based predictions. For the DrugBank dataset, these probabilities are 0.61, 0.50 and 0.50, respectively.

It is also interesting to analyze the sensitivity-specificity curves (Fig. 3C–F). Consider first the results for the Drugs.com database. For the most pressing case of uncovering previously unreported major drug interactions (Fig. 3C), we find that at 95% sensitivity, the stochastic block model has a specificity of 62%, that is, that we could have built, in 2010, a list of potential interactions containing 95% of the interactions that were actually added to the database, and excluding 62% of those that were never added. Conversely, at 95% specificity we obtain a sensitivity of 45%, that is, a list containing only 5% of the interactions that were never added to the network would have included close to half of all the interactions that were actually added to the database. While results for spurious interactions and for the DrugBank dataset are more modest, our method, unlike the neighbor-based or the Prism-based algorithms, has significant predictive power in all the cases we study.

## Application to drug discovery

Finally, we demonstrate that our algorithm can be used to discover interactions of novel drugs as part of the drug discovery process. In particular, consider a lab that has developed a new drug $D$ which is known to have a harmful interaction with another drug $H_1$. Ideally, the lab wants to identify all other drugs $\{H_2, H_3, \ldots H_m\}$ that also have harmful interactions with $D$. Since in principle there are as many potential interactions as drugs in the market (more than 1,000, according to the Drugbank and Drugs.com databases), it would be extremely costly to test all possible interactions experimentally. Considering that the typical drug interacts with approximately 20–40 other drugs (in DrugBank and Drugs.com, respectively), random testing for interactions would require 35–55 experiments to uncover a single harmful interaction.

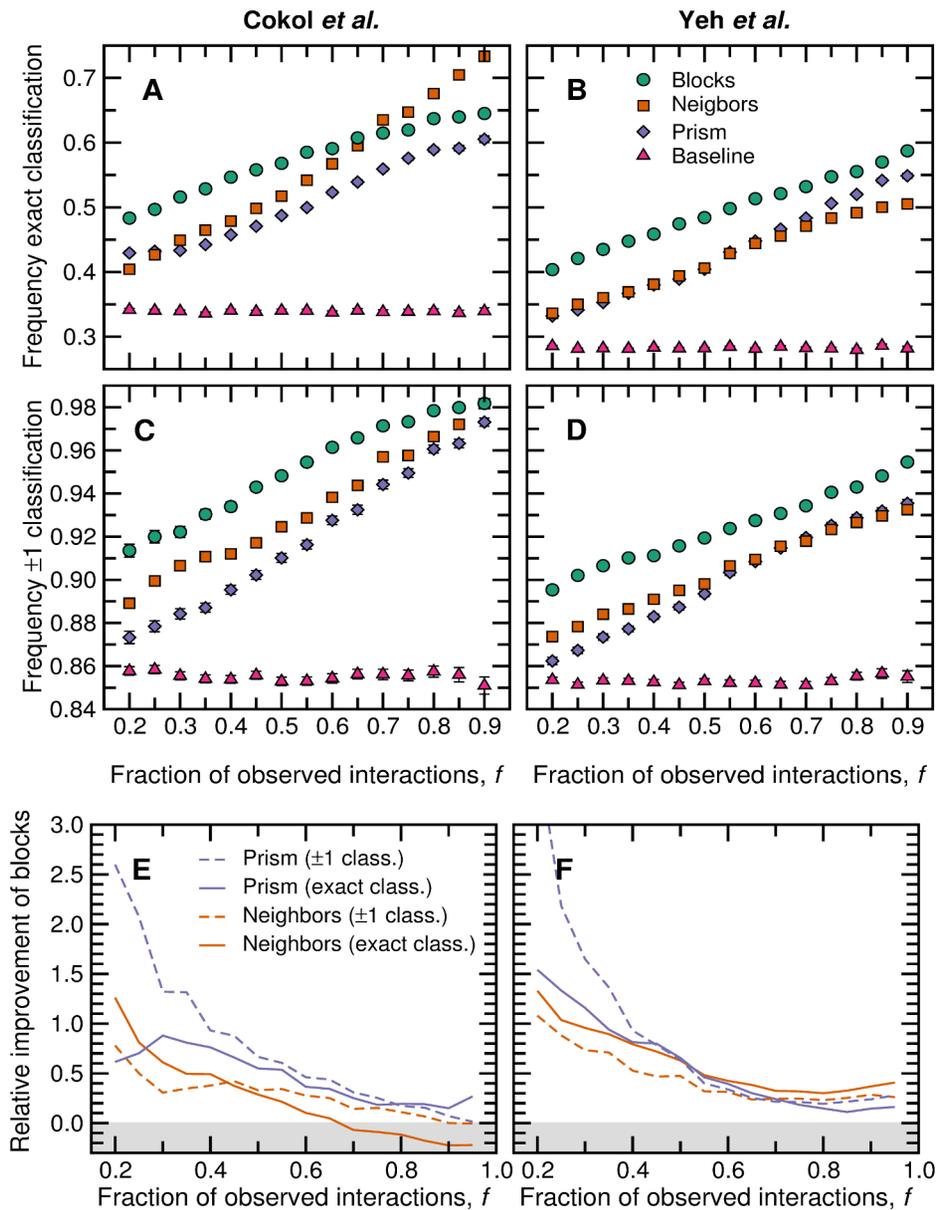

**Figure 2. Performance of drug interaction inference methods on exhaustive pair interaction data.** We test the algorithms against results of two experiments in which all pairwise interactions between a small set of drugs were tested: [28] (**A**, **C** and **E**; interactions are synergistic, additive or antagonistic) and [20] (**B**, **D** and **F**; interactions are synergistic, additive, antagonistic or suppressing). We simulate situations in which only a fraction $f$ of all interactions are observed, and then try to predict the unobserved interactions (repeated random sub-sampling validation). In each case, we measure the fraction of predictions that are exactly correct (**A** and **B**), as well as the fraction of predictions that deviate from the experimental observation by at most *one level* (**C** and **D**). For example, miss-predicting a synergistic interaction as additive is considered correct by the $\pm 1$ classification metric, but miss-predicting a synergistic interactions as antagonistic or suppressing (or vice versa), or an additive one as suppressing (or vice versa) is considered incorrect. Error bars indicate the standard error of the mean and are usually smaller than the symbols. (**E** and **F**) Relative improvement of the stochastic block model predictions over the neighbor-based predictions. If $c$ is the frequency of correct classification, we define the relative improvement as $(c_{SBM} - c_X)/(c_X - c_B)$, where SBM and B stand for stochastic block model and baseline, respectively, and X stands for any other approach (neighbor-based or Prism-based).
doi:10.1371/journal.pcbi.1003374.g002

Lacking any knowledge about $D$ (other than its interaction with $H_1$), our algorithm can guide experiments by identifying those drugs that are most likely to interact with $D$. In particular, we could use the stochastic block model inference approach to predict the most likely interaction, test it in the lab, and iterate the process adding, at each iteration, whatever interaction information the lab assay gave.

To test whether such an approach would work in practice, we have simulated the discovery of two drugs whose interactions are in fact known and reported in the 2012 snapshot of DrugBank—acetophenazine and cinacalcet (these drugs were selected randomly among those with 10 to 20 interactions). For each of these drugs, we proceed exactly as if no data were available in the database except for one seed interaction $H_1$, which we also choose at random. From the seed interaction and interaction data for all drugs other than $D$, we use the stochastic block model approach to infer the next most likely interaction of $D$, check if the interaction truly exist, add this information to the network, and iterate.

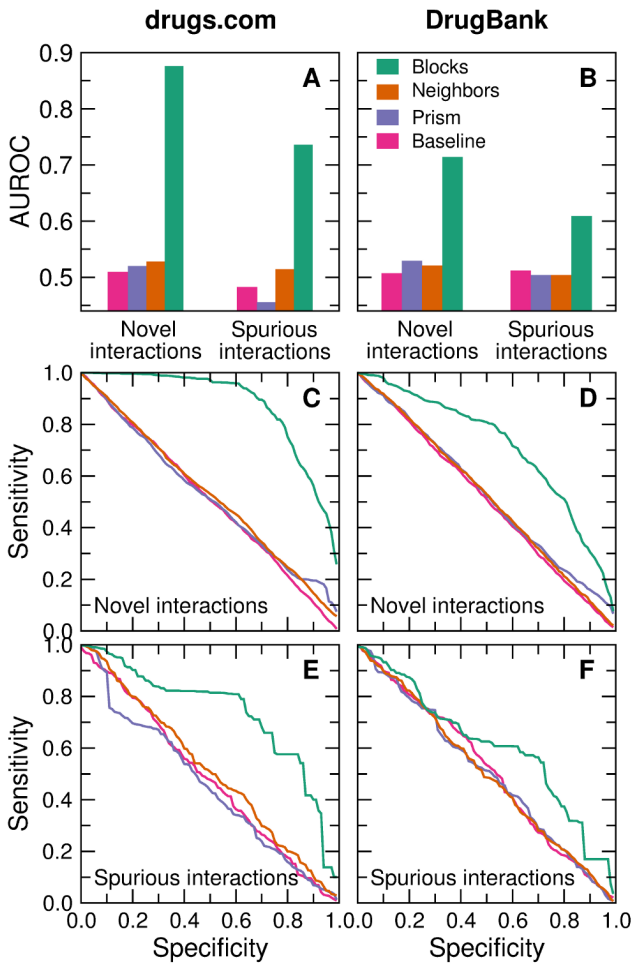

**Figure 3. Performance of drug interaction inference methods on an evolving database of major adverse drug interactions.** Left: Drugs.com database; right: DrugBank dataset. (**A–B**) Area under the receiver operating characteristic (AUROC) curve. For novel interactions the AUROC gives the probability that an interaction randomly chosen from those that were added to the first snapshot has a higher score than one randomly chosen from the set of interactions that were never added to the network. Similarly, for spurious interactions the AUROC gives the probability that an interaction randomly chosen from those that were removed from the first snapshot has a lower score than one randomly chosen from the set of interactions that were not removed from the network. (**C–F**) Sensitivity-specificity curves for novel (**C–D**) and spurious interactions (**E–F**). Sensitivity is defined as the ratio of true positives to all real positives (true positives plus false negatives). Specificity is defined as the ratio of true negatives to all real negatives (true negatives plus false positives).
doi:10.1371/journal.pcbi.1003374.g003

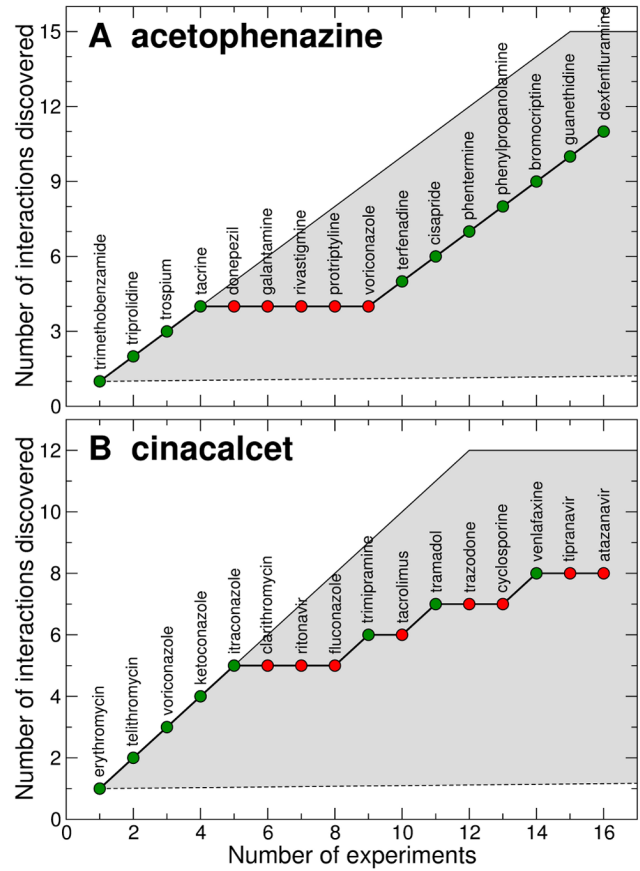

**Figure 4. Inference of drug interactions as part of the process of drug discovery and development.** For each of the two drugs ((**A**) acetophenazine and (**B**) cinacalcet) we simulate an iterative process in which a plausible interaction is suggested by the stochastic block model inference approach, the interaction is tested, and information is added to the network of known drug-drug interactions. The graphs display the number of true interactions discovered as a function of the number of experiments carried out. Green dots represent true interactions, whereas red dots represent drugs that were suggested as interaction candidates but turned out not to interact with the target drug. For acetophenazine, the 16 iterations we carry out are enough to discover 11 of the 15 interactions that are reported in DrugBank. For cinacalcet, we are able to uncover 8 of the 12 reported interactions. The gray region indicates the feasible region of discovery. Its upper bound corresponds to discovering all interactions without ever testing a drug that does not interact with the target drug; the lower bound corresponds to randomly exploring all possible interactions. In the lower bound, it takes around 100 experiments to uncover each interaction.
doi:10.1371/journal.pcbi.1003374.g004

As we show in Fig. 4, the results are very promising. For acetophenazine, the 16 iterations we carry out are enough to discover 11 of the 15 interactions that are reported in DrugBank. For cinacalcet, we are able to uncover 8 of the 12 reported interactions. As mentioned above, these numbers need to be compared with the approximately 55 experiments that would be necessary to uncover a single interaction without any guidance.

## Discussion

There is a pressing need to elucidate and understand interactions between drugs. With thousands of drugs in the market, and hundreds or thousands being tested and developed, it is clear that we cannot rely only on experimental assays to uncover interactions. Therefore, we need to develop computational data-mining methods to guide experimental analysis.

There are many possible approaches to predict drug interactions computationally. One is to mine patient data that are collected as part of post-marketing surveillance. However, this approach is problematic because of confounding factors that may not be properly accounted for in existing reporting systems [32]. Another approach is to use models based on molecular and pharmacological data [15].

Our approach is complementary to these efforts, and exploits the information that is encoded in the network of known drug interactions—since known interactions are the result of certain (known or unknown) "pharmacological rules", we can infer "rules" from known interactions and then use the inferred "rules" to, in turn, predict unreported interactions (as we show in the Supporting Text S1 and Fig. S2, the inferred "rules" correlate strongly with drug structure, category and target). Although the network approach has been frequently invoked as a new paradigm in pharmacology [7,8] and there are large-scale databases that compile and report drug interactions [10,30], this is, to the best of our knowledge, the first attempt to use network inference to predict drug interactions.

The network inference algorithm we have presented is very abstract and does not take into consideration any information other than reported interactions. It may be necessary in the future to complement the method with chemical, biological and/or pharmacological information. However, one advantage of our abstract approach is that, precisely because it is abstract, it can be easily extended to other kinds of pharmacological interaction data that can be represented as networks. For example, it is straightforward to extend our approach to predict associations between drugs and adverse side effects from *pharmacosafety networks* [33], protein- and target-drug interactions [34,35], or associations between drugs and therapies [15] and drugs and diseases [36], which may help to guide drug repositioning. Our approach can even be used to predict gene-disease associations [37] and, therefore, to uncover novel targets.

Another interesting extension of our approach is to predict multidrug interactions (that is, interactions between groups of three or more drugs), which are relevant to cancer treatment among others. Although it seems that interactions between pairs of interactions may be enough to describe higher-order interactions [16], within our framework tertiary interactions could also be modeled using three-dimensional stochastic block models in which the probability $Q(\sigma_A, \sigma_B, \sigma_C)$ that three drugs $A$, $B$ and $C$ interact depends only on the groups $\sigma_A$, $\sigma_B$ and $\sigma_C$ to which they belong. The generalization to interactions between any number of drugs is straightforward. All in all, we think that our approach opens the door to new ways of looking at and making predictions from pharmacological networks.

## Methods

### Dataset collection

For the Yeh *et al.* dataset, we collected the data on pairwise combinations of 21 antibiotics from Figs. 3 and 4a of [20]. For the Cokol *et al.* dataset, we collected the data on pairwise combinations of 13 anti-fungal drugs from Fig. 3 of [28].

For the Drugs.com dataset, we collected all drug interactions that were listed in the website, starting from a small set of highly connected seed drugs. Drugs that are not connected, directly or indirectly, to the seed drugs are not included in our analysis. We limited our searches to "generic drugs" (which include common combinations of generic drugs such as acetaminophen/hydrocodone) and to "major interactions." We consider two snapshots of the database from May 10, 2010, and February 22, 2012.

Finally, for the DrugBank dataset, we downloaded two snapshots of the complete database, corresponding to January 2009 and April 2012, from http://www.drugbank.ca/downloads [29,30].

## Estimation of link type probability using stochastic block models

The fundamental assumption of our approach is that the structure of the drug interaction network can be satisfactorily accounted for by a model $M$, which is unknown but belongs to a family $\mathcal{M}$ of models, that is, a group of models that can be parametrized in some consistent way. Then, the probability that $r_{ij} = R$ given the observed network $N^O$ is [19]

$$p(r_{ij} = R|N^O) = \int_{\mathcal{M}} dM\, p(r_{ij} = R|M) p(M|N^O), \qquad (1)$$

To estimate this integral we rewrite it, using Bayes theorem, as [19,38]

$$p(r_{ij} = R|N^O) = \frac{\int_{\mathcal{M}} dM\, p(r_{ij} = R|M) p(N^O|M) p(M)}{\int_{\mathcal{M}} dM\, p(N^O|M) p(M)}. \qquad (2)$$

Here, $p(N^O|M)$ is the probability of the observed interactions given a model and $p(M)$ is the *a priori* probability of a model, which we assume to be model-independent $p(M) = \text{const.}$

For the family of stochastic block models, each model $M = (P, \mathbf{Q}^1, \ldots, \mathbf{Q}^K)$ is completely determined by a partition $P$ of drugs into groups and the group-to-group interaction probability matrices $\mathbf{Q}^R$. Here, $K$ is the total number of interaction types (for example, if interactions can be synergistic, additive or antagonistic, then $K = 3$) and, for a given partition $P$, the matrix element $Q^R(\alpha, \beta)$ is the probability that a drug in group $\alpha$ and a drug in group $\beta$ interact with each other (these matrices verify that $\sum_r Q^r(\alpha, \beta) = 1$ for all pairs of groups $(\alpha, \beta)$). Thus, if $i$ belongs to group $\sigma_i$ and $j$ to group $\sigma_j$ we have that [38]

$$p(r_{ij} = R|M) = Q^R(\sigma_i, \sigma_j) ; \qquad (3)$$

and

$$p(N^O|M) = \prod_{\alpha \leq \beta} \prod_r Q^r(\alpha, \beta)^{n^r(\alpha, \beta)}, \qquad (4)$$

where $n^r(\alpha, \beta)$ is the number of interactions of type $r$ between drug groups $\alpha$ and $\beta$.

The integral over all models in $\mathcal{M}$ can be separated into a sum over all possible partitions of the drugs into groups, and an integral over all possible values of each $Q^r(\alpha, \beta)$. Using this together with Eqs. (2), (3) and (4), and under the assumption of no prior knowledge about the models ($p(M) = \text{const.}$), we have

$$p(r_{ij} = R|R^O) =$$
$$\frac{1}{Z} \sum_P \int_S d\mathbf{Q}^1 \ldots \int_S d\mathbf{Q}^K \, Q^R(\sigma_i, \sigma_j) \prod_{\alpha \leq \beta} \prod_r Q^r(\alpha, \beta)^{n^r(\alpha, \beta)}; \qquad (5)$$

where the integral is over all $Q^r(\alpha, \beta)$ within the subspace $S$ that satisfies the normalization constraints $\sum_r Q^r(\alpha, \beta) = 1$, and $Z$ is the normalizing constant (or partition function). These integrals factorize into terms corresponding to all pairs $(\alpha, \beta)$ [38], each with the general form

$$\int_0^1 dQ^1 (Q^1)^{n^1} \int_0^{1-Q^1} dQ^2 (Q^2)^{n^2} \cdots \int_0^{1-Q^1-\cdots-Q^{K-2}}$$

$$dQ^{K-1}(Q^{K-1})^{n^{K-1}}(1-Q^1-\cdots-Q^{K-1})^{n^K}$$

$$= \frac{n^1! n^2! \cdots n^K!}{(n^1+n^2+\cdots+n^K+K-1)!}.$$

Using these expressions in Eq. (5), one obtains

$$p(r_{ij} = R | N^O) = \frac{1}{Z} \sum_P \left( \frac{n^R(\sigma_i, \sigma_j)+1}{n(\sigma_i, \sigma_j)+K} \right) \exp(-H(P)), \quad (6)$$

where the sum is over all partitions of the drugs, $n(\sigma_i, \sigma_j) = \sum_r n^r(\sigma_i, \sigma_j)$ is the total number of known interactions between groups $\sigma_i$ and $\sigma_j$, and $H(P)$ is a function that depends on the partition only

$$H(P) = \sum_{\alpha \le \beta} \left[ \ln(n(\alpha, \beta)+K-1)! - \sum_{r=1}^K \ln(n^r(\alpha, \beta))! \right]. \quad (7)$$

This sum can be estimated using the Metropolis algorithm [19,39] as detailed next.

## Implementation details

The sum in Eq. (6) cannot be computed exactly because the number of possible partitions is combinatorially large, but can be estimated using the Metropolis algorithm [19,39]. This amounts to generating a sequence of partitions in the following way. From the current partition $P^0$, select a random drug and move it to a random new group giving a new partition $P^1$. If $H(P^1) < H(P^0)$, always accept the move; otherwise, accept the move only with probability $e^{H(P^0)-H(P^1)}$.

By doing this, one gets a sequence of partitions $\{P^i\}$ such that [39]

$$p(r_{ij} = R | N^O) \approx \frac{1}{N} \sum_{P \in \{P^i\}} \frac{n^R(\sigma_i, \sigma_j)+1}{n(\sigma_i, \sigma_j)+K}, \quad (8)$$

where $N$ is the number of samples in $\{P^i\}$.

In practice, it is useful to "thin" the sample $\{P^i\}$, that is, to consider only a small fraction of evenly spaced partitions so as to avoid the computational cost of sampling very similar partitions which provide very little additional information. Moreover, one needs to make sure that sampling starts only when the sampler is "thermalized", that is, when sampled partitions are drawn from the desired probability distribution (which in our case is given by $e^{-H(P)}/Z$). Our implementation automatically determines a reasonable thinning of the sample, and only starts sampling when certain thermalization conditions are met. Therefore, the whole process is completely unsupervised. The source code of our implementation of the algorithm is publicly available from http://seeslab.info/downloads/drugraph/ and http://github.com/seeslab/drugraph.

As often happens in Metropolis sampling, in general it is better to run many short independent sampling processes that a single very long sampler. Results reported here are obtained using 50 independent sampling processes of 200 (conveniently thinned) partitions each. These sampling processes can be run in parallel, taking on the order of 1–2 days to complete on high-end CPUs for the largest network considered here (with over 1,500 drugs). Sampling an equivalent 10,000 partitions with a single run can take 2–3 weeks.

## Prism-based prediction of interactions

The Prism algorithm [27] was originally developed to identify groups of genes that interact monochromatically, that is, that have the same type of interactions with genes in other groups. Yeh and coworkers then introduced Prism II [20] to identify groups of drugs relaxing the requirement for perfect monochromaticity.

Our implementation of Prism II is as follows. Each drug is initially placed in a group by itself. Then, groups are sequentially merged until all drugs belong to a single group. At each step, we merge the two groups with the smallest "distance" to each other. The distance $F_{\alpha\beta}$ between groups $\alpha$ and $\beta$ is

$$F_{\alpha\beta} = \min_{i \in \alpha, j \in \beta} \{d_{ij}\} - T \Delta S_{\alpha\beta}. \quad (9)$$

Here, the normalized drug-drug distance $d_{ij}$ between drugs $i$ and $j$ is

$$d_{ij} = \frac{1}{K-1} \sqrt{\frac{\sum_k (r_{ik} - r_{jk})^2}{N_{ij}}}, \quad (10)$$

with $N_{ij}$ the number of interactions reported for both $i$ and $j$. The change of monochromaticity entropy $\Delta S_{\alpha\beta}$ is

$$\Delta S_{\alpha\beta} = S(\mathbf{m}_{\alpha\beta} + \mathbf{m}_{\alpha\alpha} + \mathbf{m}_{\beta\beta})$$
$$- S(\mathbf{m}_{\alpha\beta}) - S(\mathbf{m}_{\alpha\alpha}) - S(\mathbf{m}_{\beta\beta})$$
$$+ \sum_{\sigma \ne \alpha\beta} \left[ S(\mathbf{m}_{\alpha\sigma} + \mathbf{m}_{\beta\sigma}) - S(\mathbf{m}_{\alpha\sigma}) - S(\mathbf{m}_{\beta\sigma}) \right] \quad (11)$$

where $\mathbf{m}_{\alpha\beta} = (m_{\alpha\beta}^-, m_{\alpha\beta}^+)$ is a vector with the number of synergistic ($-$) and antagonistic ($+$) interactions between groups $\alpha$ and $\beta$, and

$$S(\mathbf{m}) = \frac{m^+ + m^-}{M^+ + M^-} (p^+ \log p^+ + p^- \log p^-) \quad (12)$$

with $p^{+/-} = m^{+/-}/(m^+ + m^-)$ and $M^{+/-} = \sum_{\alpha\beta} m_{\alpha\beta}^{+/-}$.

By itself, the Prism II algorithm returns a tree of nested drug groupings. To make interaction predictions, we need to: (i) set the free parameter $T$; (ii) cut the tree at a certain level to get a single partition of the drugs into groups (a process that needs to be unsupervised); and (iii) given those groups, determine the probability of each type of interaction. To cut the tree, we choose the partition with the smallest number of groups among those with total monochromaticity entropy $S = \sum_{\alpha,\beta} S(\mathbf{m}_{\alpha\beta})$ that satisfies $S < 0.10 S_{max}$, where $S_{max}$ is the partition that corresponds to putting all drugs in a single group. Additionally, we set $T = 10$ to get results consistent with those reported in Ref. [20] (we also checked that these parameters yield good results for the Cokol dataset, and that the results do not improve using other values of $T$; see Supporting Text S1 and Fig. S1).

Finally, once the groups are defined, we estimate the probability $p^{Prism}(r_{ij} = R | N^O)$ as

$$p^{\mathrm{Prism}}(r_{ij} = R|N^O) = \frac{n^R(\sigma_i, \sigma_j)}{n(\sigma_i, \sigma_j)}, \qquad (13)$$

where $n^R(\sigma_i, \sigma_j)$ and $n(\sigma_i, \sigma_j)$ are defined as above.

With our implementation, the Prism-based algorithm takes 1–2 days on high-end CPUs to generate interaction predictions for the large networks considered here (with over 1,000 drugs).

### Neighbor-based prediction of interactions

Given a network of drug interactions, we define the interaction similarity $s_{ik}$ between drugs $i$ and $k$ as the fraction of interactions with other drugs that are equal for $i$ and $k$, over the total number of interactions that are reported for both drugs. In particular $s_{ii} = 1$, and $s_{ik} = 0$ if two drugs do not have any equal interaction with others.

To predict the interaction $r_{ij}$ between drugs $i$ and $j$, we order all possible drug pairs $(k_i, k_j)$ by decreasing value of the product of similarities to the query drugs $s_{ik_i} \times s_{jk_j}$. We then select the pair $(k_i = i', k_j = j')$ with the highest product for which the interaction $r_{i'j'}$ is known, and use that value as our prediction for $r_{ij}$. Note that we may have $i' = i$, that is, we may use the known interaction between $i$ and a drug $j'$ that is very similar to $j$ to predict $r_{ij}$.

## Supporting Information

**Figure S1** The accuracy of the Prism-based method, as measured by the AUROC, does not improve consistently and significantly by choosing values of $T$ other than $T = 10$, as used in the main text.
(TIFF)

**Figure S2** Drug groups and drug mechanisms of action from stochastic block models. For each drug pair in the Cokol et al. dataset (**A–B**), the Yeh et al. dataset (**C–D**) and the DrugBank 2012 snapshot (**E**), we calculate the probability that any two drugs belong to the same drug group (see Section 2). We call this probability the co-classification probability. (**A**) and (**C**) The matrix

of co-classification probabilities for the Cokol et al. dataset (**A**), and Yeh et al. dataset (**C**), ordered so that large co-classification probabilities appear close to the diagonal [40]. Dashed lines are a guide to the eye. The mechanism of action of each drug is indicated by color bars on top of drug abbreviations ((**A**) Cyan: ergosterol metabolism; dark red: acting on serine/threonine; other drugs were intentionally selected with different targets and mechanisms of action. (**C**) Dark red: protein synthesis, 30S; cyan: protein synthesis, 50S; red: folic acid biosynthesis; pink: DNA gyrase; dark blue: cell wall; yellow: aminoglycoside, protein synthesis, 30S. Co-classification boxes correspond, to a large extent, to mechanisms of action. (**B**) and (**D**) The reported drug interactions show clear patterns once they are ordered according to the co-classification probability. For example in the Yeh et al. dataset, most interactions between the group {AMK,STR,TOB} and the group {TMP,SLF,NIT,CPR,LOM} are synergistic. (**E**) We use information in DrugBank to analyze the overlap (or functional similarity) in substructure, category and target between pairs of drugs (see Section 3). We plot these quantities as a function of the co-classification probability of the corresponding drug pairs (we average over drug pairs with similar co-classification probability; error bars represent the standard error of the mean and are generally smaller than the symbols). Drugs with higher co-classification probability are significantly more likely to share substructures, categories and targets.
(TIFF)

**Text S1** Sensitivity analysis for the Prism-based algorithm. Discussion on drug groups and drug mechanisms. Discussion on the estimation of the co-classification probability using stochastic block models. Analysis of drug similarities for drugs in the DrugBank.
(PDF)

## Author Contributions

Conceived and designed the experiments: RG MSP. Performed the experiments: RG. Analyzed the data: RG MSP. Contributed reagents/materials/analysis tools: RG MSP. Wrote the paper: RG MSP.